\documentclass[aps,prl,showpacs,twocolumn,superscriptaddress,preprintnumbers,nofootinbib]{revtex4-1}

\pdfoutput=1

\usepackage{amsmath, amsfonts, amsthm, amssymb, graphicx}

\usepackage{amsmath}
\usepackage{amsfonts}
\usepackage{amssymb}
\usepackage{subeqnarray}
\usepackage{latexsym}
\usepackage[latin1]{inputenc}
\usepackage{graphicx}

\usepackage[english]{babel}

\usepackage{hyperref}
\hypersetup{
pdfstartview=FitV,
colorlinks=true,
linkcolor=blue,
citecolor=blue,
filecolor=blue,
urlcolor=black}

\allowdisplaybreaks[3]

\makeatletter
\@addtoreset{equation}{section}
\makeatother

\newcommand{\beq}{\begin{eqnarray}}
\newcommand{\eeq}{\end{eqnarray}}

\newcommand{\bea}{\begin{eqnarray}}
\newcommand{\eea}{\end{eqnarray}}
\def\non{\nonumber}

\def\p{\psi}

\def\non{\nonumber}

\def\bx{\bf{x}}

\begin{document}

\rightline{DAMTP-2017-18}
\vspace{0.5cm}

\title{Exact Gravitational Wave Signatures from Colliding Extreme Black Holes}

\author{Joan Camps}
\affiliation{Department of Physics and Astronomy, \\
 University College London, Gower Street, London WC1E 6BT, UK}

\author{Shahar Hadar}
\affiliation{Department of Applied Mathematics and Theoretical Physics, \\
 University of Cambridge, Wilberforce Road, Cambridge CB3 0WA, UK}

\author{Nicholas S. Manton}
\affiliation{Department of Applied Mathematics and Theoretical Physics, \\
 University of Cambridge, Wilberforce Road, Cambridge CB3 0WA, UK}

\begin{abstract}
The low-energy dynamics of any system admitting a continuum of static configurations is approximated by slow motion in moduli (configuration) space. Here, following Ferrell and Eardley, this moduli space approximation is utilized to study collisions of two maximally charged Reissner--Nordstr{\"o}m black holes of arbitrary masses, and to compute analytically the gravitational radiation generated by their scattering or coalescence. The motion remains slow even though the fields are strong, and the leading radiation is quadrupolar. A simple expression for the gravitational waveform is derived and compared at early and late times to expectations.
\end{abstract}

\maketitle

\section{Introduction} \label{Introduction}

The two-body problem in General Relativity has seen extensive study in the last few decades. This research program is motivated first and foremost by the need for accurate predictions of gravitational wave (GW) signatures to be measured by observatories such as LIGO \cite{LIGO}, LISA \cite{LISA}, and others. Nowadays, after the remarkable first direct detections \cite{Abbott:2016blz}, the field is entering an exciting stage in which theory and experiment interact, and the demand for precision predictions is enhanced. Theoretically, several different approaches have been developed to tackle the challenge. Analytical progress has been made in two limits: the Post-Newtonian approximation (PN, c.f.~\cite{Blanchet:2006zz} for a review; see also \cite{Porto:2016pyg}), in which fields are weak; and the extreme mass ratio limit (c.f.~\cite{Poisson:2011nh}), in which one of the binary's constituents is much smaller than the other. Numerically, since the 2005 breakthrough \cite{Pretorius:2005gq}, it has been possible to simulate merging binary systems fully nonlinearly. All existing approaches coalesce in the Effective One Body (EOB) framework \cite{Buonanno:1998gg}, which describes evolution in terms of geodesic motion on an effective spacetime geometry which is calibrated by PN, extreme mass ratio, and numerical input.

In this paper we use a different approximation scheme in order to compute analytically the GW signatures from scattering and merging black holes (BHs) in the strong-field regime and for any mass ratio: the \emph{moduli space approximation} (MSA), in which the system evolves adiabatically through a series of approximately static configurations. Only in special cases does a multiple BH system admit such a nontrivial space (\emph{moduli space}) of stationary, degenerate configurations; indeed, we study a special binary composed of two non-rotating, maximally charged (extreme Reissner--Nordstr{\"o}m (ERN)) BHs, between which the static gravitational and electric forces cancel. Even though this specific system is not expected to be relevant astrophysically, our approach provides novel insight into strong-field gravity. To the best of our knowledge, this is the first time that GWs emitted from the high-curvature region of a binary BH spacetime of generic mass ratio are computed analytically.

The MSA was developed to deal with multi-soliton dynamics in field theories \cite{Manton:1981mp}. It applies to several types of topological solitons where the static forces cancel out, and where the leading forces are $\mathcal{O}(v^2)$, where $v$ is a typical relative speed. For example, it applies to BPS magnetic monopoles and gauged vortices, where the static magnetic and Higgs scalar forces
cancel. In the MSA, soliton motion at non-relativistic speeds is modelled
by a geodesic motion through the moduli space of static multi-soliton
solutions, and this idea has been verified by rigorous analysis in some cases \cite{Stuart:1994tc}.  The MSA is applicable to certain types of gravitating solitons, e.g. Kaluza--Klein monopoles \cite{Ruback:1986ag} and higher-dimensional supersymmetric BHs \cite{Michelson:1999dx}, and was first applied to the ERN BHs considered in this paper by Gibbons and Ruback, who studied well-separated objects \cite{Gibbons:1986cp}, and by Ferrell and Eardley, who found the complete two-body moduli space geometry \cite{Ferrell:1987gf,FE2}. Here it is shown how to extract GW signatures, in closed form, from Ferrell and Eardley's pioneering analysis of the BHs' motion.

It is possible in principle to calculate the metric on moduli space directly from the kinetic energy expression in the field theory, as for example in \cite{Strachan:1992fb}, but alternatively, it may be found by
sophisticated geometrical arguments \cite{Atiyah:1988jp}, or by numerical
computation \cite{Samols:1991ne}. For well-separated solitons, the asymptotic moduli space metric can be found by treating the solitons as point-like particles and calculating their interactions, carefully including the
velocity-dependent parts of the forces \cite{Gibbons:1995yw,Gibbons:1986cp}, which do not cancel.

Radiation associated with multi-soliton dynamics has received some attention previously. Manton and Samols calculated the electromagnetic (EM) and scalar radiation during a head-on collision of two SU(2) monopoles \cite{Manton:1988bn}.
One can calculate the time-dependent, asymptotic quadrupole fields exactly, and hence find the leading-order radiation. A similar approach is adopted here, but in a gravitational context.

The GW signatures we compute should be useful for comparison with other approaches to the relativistic two-body problem, and will hopefully provide new insight. First, it will be interesting to compare with numerical simulations. In \cite{Zilhao:2012gp} the numerical study of charged BH collisions was initiated; there, the BHs were initially at rest, and the generalization to (near-extreme) BHs with nonzero initial velocities could be compared with our results. Second, and a touch more speculatively, it would be interesting to try and make contact with the EOB scheme. It is intriguing that the natural description of our ERN binary system is by geodesic motion in an effective geometry, just as suggested by the EOB approach. Finally, it may be interesting to check whether any of the lessons of this paper apply to systems of astrophysically relevant, electrically neutral, near-extremal \emph{rotating (Kerr)} BHs (see \cite{Hadar:2016vmk} for recent progress in the extreme mass ratio limit).

\section{Slowly moving extreme black holes} \label{FerrelEardley}
In Einstein--Maxwell theory, BHs can carry a maximum amount of charge per unit mass, as $|Q|\leq M$. At extremality, charged BHs with $|Q_a|=M_a$, and all the charges of equal sign, coexist in static equilibrium due to the exact cancellation of electrostatic repulsion  and gravitational attraction. These configurations (with all charges positive) have a remarkably simple description in the Majumdar--Papapetrou form \cite{Majumdar:1947eu,Papapetrou:1947ib}, with metric and EM 1-form potential\footnote{in natural units $c=G_N=1$}
\bea
ds^2 &=& - \p^{-2} \,  dt^2 + \p^2 \,  d{\bx}\cdot d{\bx} \, , \non \\
A &=& - (1-\p^{-1}) dt\, ,
\label{eq:MPansatz}
\eea
where
\bea
\p = 1 + \sum_{a=1}^N\frac{M_a}{\left| {\bx} - {\bx}_a\right|} \, .
\label{eq:psiStationary}
\eea

Since there is no net interaction, the 3-vector positions of the BHs ${\bx}_a$ can be chosen freely: they are the \emph{moduli} of these configurations. When the BH motion is slow, i.e.~$v \ll 1$ where $v \sim | \dot{{\bx}}_a |$ is a typical speed, one can consider the moduli space, or adiabatic, approximation, in which snapshots of the system are well approximated by static fields. The BHs do not, however, move freely along straight lines, as there are higher-order corrections to the Coulombic forces, leading to magnetic and gravitomagnetic effects at $\mathcal{O}(v^2)$ that do not cancel.

The MSA is constructed systematically by promoting ${\bx}_a\rightarrow {\bx}_a(t)$ in eqs.~\eqref{eq:MPansatz}--\eqref{eq:psiStationary}, and adding to the ansatz  $\mathcal{O}(v)$ corrections to solve the field equations to first order in velocities \cite{Gibbons:1986cp,Ferrell:1987gf,FE2}.\footnote{Since there is no dynamics at leading order, the relevant equations are the constraints.} Evaluating on-shell the Einstein--Maxwell action on the $\mathcal{O}(v)$ solutions produces an effective action for the moduli:\footnote{See \cite{Andrade:2015qea} for a critical discussion of this approximation.}
\beq
S_{\textrm{on-shell}}=\int dt\, g_{AB}({\bx}_1,\ldots,{\bx}_N)\, \dot{x}^A(t)\dot{x}^B(t)\, ,
\label{eq:effectiveAction}
\eeq
where upper-case indices run over all components of the moduli: $A=1,\dots,3N$. The effective action \eqref{eq:effectiveAction}, which is purely kinetic, determines the system's evolution at leading order.

Importantly, the Lagrangian in \eqref{eq:effectiveAction} is seen to be of $\mathcal{O}(v^2)$. This is in contrast to the more familiar PN approximation in which the $\mathcal{O}(v^2)$ terms, including the so-called Einstein--Infeld--Hoffman Lagrangian \cite{Einstein:1938yz} (see also \cite{Blanchet:2006zz}), are combined with the leading order, Newtonian potential. For the EBHs we are considering, the $\mathcal{O}(v^0)$ Lagrangian is the (constant) total mass, promoting \eqref{eq:effectiveAction} to be the leading non-trivial term in the effective action. This is, in fact, the characteristic feature enabling the MSA.

The interpretation of the action \eqref{eq:effectiveAction}, as realised in \cite{Manton:1981mp}, is that $g_{AB}$ is a metric on the purely spatial, $3N$-dimensional moduli space. The extrema of \eqref{eq:effectiveAction} describe motion along geodesics at constant speed in this geometry. Since the Newtonian potential is absent and kinetic energy is conserved, the motion can remain slow and the MSA stays controlled, even though the fields are strong when $|{\bx}_a-{\bx}_b| \lesssim \textrm{min}(M_a, M_b)$.

As radiation losses are of higher orders in velocities, the total energy and momentum are constants of motion through second order in the MSA. The center of mass thus moves freely, so its coordinates span a flat $\mathbb{R}^3$ factor in moduli space. For the two-body problem, $N=2$, the remaining three coordinates are conveniently taken to be the relative position, ${\bx}_*\equiv{\bx}_1-{\bx}_2$. As Ferrell and Eardley showed \cite{Ferrell:1987gf}, the non-trivial factor in the moduli space metric reads:
\beq
\mu \, \gamma \, d{\bx}_* \cdot d{\bx}_* \, ,
\label{eq:moduliGeometry}
\eeq
where
\beq
\gamma = \left[\left(1+\frac{M}{r_*}\right)^3-\frac{2\mu\, M^2}{r_*^3}\right]\,,
\eeq
with $M\equiv M_1+M_2$ and $\mu\equiv\frac{M_1 M_2}{M_1+M_2}$ the total and reduced masses, and $r_*\equiv|{\bx}_*|$. This geometry is spherically symmetric, and its equatorial sections interpolate between flat space as $r_*\rightarrow\infty$ and an infinite flat cone with deficit angle $\pi$ as $r_*\rightarrow 0$, as can be seen by changing coordinates to $r_*\propto \rho^{-2}$ and expanding for large $\rho$.

Spherical symmetry confines the geodesics of \eqref{eq:moduliGeometry} to equatorial planes. Such geodesics solve the radial ODE
\beq
\left(\frac{dr_*}{dt}\right)^2+\frac{v_\infty^2}{\gamma}\left[\frac{b^2}{\gamma\, r_*^2}-1\right]=0\,\, ,
\eeq
where $v_\infty$ is the relative speed at infinite separation. $b$ is the impact parameter, and sets the angular momentum to $J=\mu\, b\, v_\infty$. $b_{\textrm{crit}}(\mu, M)$ divides trajectories in moduli space into coalescing ($b<b_{\textrm{crit}}$) and scattering ($b>b_{\textrm{crit}}$) trajectories.\footnote{$b_{\textrm{crit}}(\mu, M)$ is a solution to a cubic equation given in \cite{FE2}. It does not depend on $v_\infty$ because of the velocity dependence of the forces.  $v_\infty$ just sets the time scale of the motion.}

It is interesting to examine the early- and late-time limits of the trajectory. For scattering orbits both limits correspond to large separation $r_* \gg M$, while for coalescing orbits late times correspond to small separation $r_* \ll M$.
At early times, $t \to -\infty$, to next-to-leading order,
\bea
r_* = -v_\infty\,t-\frac{3}{2} M \ln (-v_\infty\, t)+ \cdots \, .
\label{eq:earlyTimes}
\eea
At late times, $t \to \infty$, scattering orbits are given by \eqref{eq:earlyTimes} with $-v_\infty$ replaced by $v_\infty$.
For coalescing orbits,
\beq
r_*= \frac{C_\infty}{(v_\infty\,t)^2}+\cdots\,\,,\qquad C_\infty\equiv4M^2(M-2\mu)\, ,
\label{eq:lateTimes}
\eeq
where the power law can be understood directly from the moduli space geometry: since the asymptotic cone at small $r_*$ is flat, motion is free, so $\rho\propto t$ and \eqref{eq:lateTimes} follows. Note that the motion extends for an infinite time because the moduli space is geodesically complete.

\section{Radiation from an extreme binary}\label{radiation}
The analysis of wave emission from slowly evolving systems is facilitated by the fact that, when $v\ll1$, the two relevant length scales become parametrically disparate. One scale is the size of the system $L \sim \max\{r_*,M\}$, defining the \emph{near/system} zone; the other scale is the wavelength of emitted radiation $\lambda \gtrsim L/v \gg L$, defining the \emph{far/radiation} zone.

In the near zone, wave propagation is approximately instantaneous, as the wavelength appears infinite; in the radiation zone, waves originate from a shrunk, point-like source with a number of radiation multipole moments. It was shown by Thorne \cite{Thorne:1980ru} that the matching between the two zones results in the radiation multipoles equalling the system multipoles, dominated by the quadrupole. The system multipoles are defined in the near zone as coefficients of a large-distance expansion from the source $r\gg r_*$, with a particular algorithm \cite{Thorne:1980ru} that applies in spite of the possibility of strong gravity when the system size is comparable to its gravitational radius $r_*\lesssim M$.

The leading gravitational wave emission is then given by the quadrupole formula
\bea
h^{TT} = \frac{2}{r} \left.\frac{d^2}{dt^2}Q^{TT}\right|_{t_\mathrm{ret}} \, ,
\label{eq:quadFormula}
\eea
where $h$ is the metric perturbation tensor describing the gravity wave, $Q$ is the near zone quadrupole moment tensor, $TT$ means the transverse-traceless projection, and where $t_{\mathrm{ret}} \equiv t-r$ is the retarded time.

The algorithm calculating the mass quadrupole moment in the near zone involves a large $r$ expansion in \emph{asymptotically Cartesian, mass-centered (ACMC) gauge}, such that the lowest power of $1/r$ accompanying the spherical harmonic $Y_{\ell}^{m}(\theta,\phi)$ is $1/r^{\ell+1}$.
The metric \eqref{eq:MPansatz} is in such gauge, and the relevant large $r$ expansion of the time-time component reads\footnote{Recall that motion happens in the equatorial plane $\theta_*=\pi/2$. $\{r, \theta, \phi\}$ are coordinates of an observer.}
\begin{widetext}
\beq
\psi^{-2}\approx
1-\frac{2M}{r}+3\frac{M^2}{r^2}-4\frac{M^3}{r^3}+\mu\frac{r_*^2}{r^3}\left[1-3\sin^2\theta\cos^2(\phi-\phi_*)\right] \, .
\label{eq:g_00_expansion}
\eeq
The angular dependence in brackets is proportional to $\left[e^{-2i\phi_*}Y^2_2-\sqrt{\frac{2}{3}}Y^0_2+e^{2i\phi_*}Y^{-2}_2\right]$, and it is the first appearance of $\ell=2$ harmonics, at order $1/r^3$. Thus, its coefficients define the system's mass quadrupole \cite{Thorne:1980ru}. Notice that this quadrupole is defined even in strong gravity, when the $M^3$ and $\mu\, r_*^2$ contributions to the $1/r^3$ tails are comparable. ACMC gauge confines non-linearities in these tails to the $\ell=0$ sector, giving an effectively linear (in the coupling constant $G_N$) $\ell=2$ quadrupole, which can then feed into the quadrupole formula \eqref{eq:quadFormula}. The $TT$ projection gives
\bea
Q^{TT} = \frac{\mu\, r_*^2}{4} \, \left[e^{-2i\phi_*}{}_{-2}Y^2_2-\sqrt{\frac{2}{3}}\,{}_{-2}Y^0_2+e^{2i\phi_*}{}_{-2}Y^{-2}_2\right] \sqrt{2}\, \hat{e}_R+\textrm{c.c.} \, ,
\label{eq:quadrupole}
\eea
where $r_*$ and $\phi_*$ depend on time, and $\textrm{c.c.}$ stands for complex conjugation. The $s=-2$ spin-weighted spherical harmonics are\footnote{conveniently normalized such that $\int {}_{s}\bar{Y}_\ell^m\, {}_{s}Y_\ell^m\, d\Omega = 4\pi/(2\ell+1)$ \label{foot:SphH}}
\bea
{}_{-2}Y^{2}_2 = \frac{e^{2i\phi}}{(1+z \bar{z})^2} \, ,
\quad
{}_{-2}Y_{2}^0 = \frac{\sqrt{6} \, \bar{z}^2 \, e^{2i\phi}}{(1+z\bar{z})^2} \, ,
\quad
{}_{-2}Y_{2}^{-2} = \frac{\bar{z}^4 \, e^{2i\phi}}{(1+z \bar{z})^2}  \, ,
\label{SWSH_-2}
\eea
\end{widetext}
where $z=\tan(\theta/2) \,  e^{i\phi}$, and the circular polarization tensors read
\bea
\sqrt{2}\,\hat{e}_R = \frac{4r^2 \, e^{-2i\phi}}{(1+z\bar{z})^2} \, dz^2 \, , \quad \hat{e}_L=\bar{\hat{e}}_R \, .
\label{circ polarization}
\eea
Eq.~\eqref{eq:quadrupole} can also be written in a linear polarization basis using standard formulas, $\sqrt{2}\,\hat{e}_{R/L} = \hat{e}_+ \pm i\, \hat{e}_{\times}$.
Plugging \eqref{eq:quadrupole} into \eqref{eq:quadFormula} yields the gravitational waves far from the source.

The gravitational waveform is accompanied by an EM wave, as there is charge acceleration. Through order $v^2$, one may expect electric dipole, magnetic dipole, and electric quadrupole radiation. However, many simplifications take place due to the equality of charge and mass. The electric dipole vanishes through first order in $v$, as the centres of mass and charge can differ only when $\mathcal{O}(v^4)$ terms in the kinetic energy are taken into account. This vanishing can be checked explicitly by expanding the electric potential $A_0$, similarly to \eqref{eq:g_00_expansion}, and this expansion also shows that the electric quadrupole is identical to the gravitational one. The magnetic dipole does not radiate either, as it is conserved to leading order (in $v$) because it is proportional to the angular momentum. The leading EM radiation is then also quadrupolar:
\beq
A^{T} = \frac{1}{2 r}\left.\frac{d^2}{dt^2}(Q\cdot n)^T\right|_{t_\mathrm{ret}} \, ,
\eeq
where $n=dr$ is the unit vector in the observer's direction, and as before $T$ projects out the transverse part of the wave. This projection gives
\begin{widetext}
\beq
(Q\cdot n)^T=-\frac{\mu\, r_*^2}{4} \, \left[e^{-2i\phi_*}{}_{-1}Y^2_2-\sqrt{\frac{2}{3}}\,{}_{-1}Y^0_2+e^{2i\phi_*}{}_{-1}Y^{-2}_2\right]\sqrt{2}\, |R\rangle+\textrm{c.c.} \, ,
\label{waveform 1 EM}
\eea
with $s=-1$ spin-weighted spherical harmonics (normalized as in footnote \ref{foot:SphH}):
\bea
{}_{-1}Y^{2}_2 = -\frac{2 z \, e^{i\phi}}{(1+z \bar{z})^2} \, ,
\quad
{}_{-1}Y_{2}^0 = \sqrt{6}\,\frac{\bar{z}(1-z\bar{z}) \, e^{i\phi}}{(1+z\bar{z})^2} \, ,
\quad
{}_{-1}Y_{2}^{-2} =  \frac{2 \bar{z}^3e^{i\phi}}{(1+z \bar{z})^2} \, ,
\label{SWSH_-1}
\eea
\end{widetext}
and the right- and left-handed unit polarization vectors are
\beq
\sqrt{2}\,|R\rangle=\frac{2r\, e^{-i\phi}}{1+z\bar{z}}\,dz\,\,,\qquad |L\rangle = \bar{|R\rangle}\,\,.
\eeq
Note that the EM and gravitational radiation fluxes have different angular distributions; for example, at the north pole the EM signal vanishes while GWs are strongest.

The simplicity of eqs.~\eqref{eq:quadFormula} and \eqref{eq:quadrupole}, and their similarity to PN formulae, may be deceptive; it is important to stress that here they describe the radiation emitted from the strong-field region of a highly curved geometry. Despite not being PN, the slow motion of this BH system with its strong fields can be found analytically, for any mass ratio, using the MSA.

\begin{figure}[h!]
 \includegraphics[width=8.5cm]{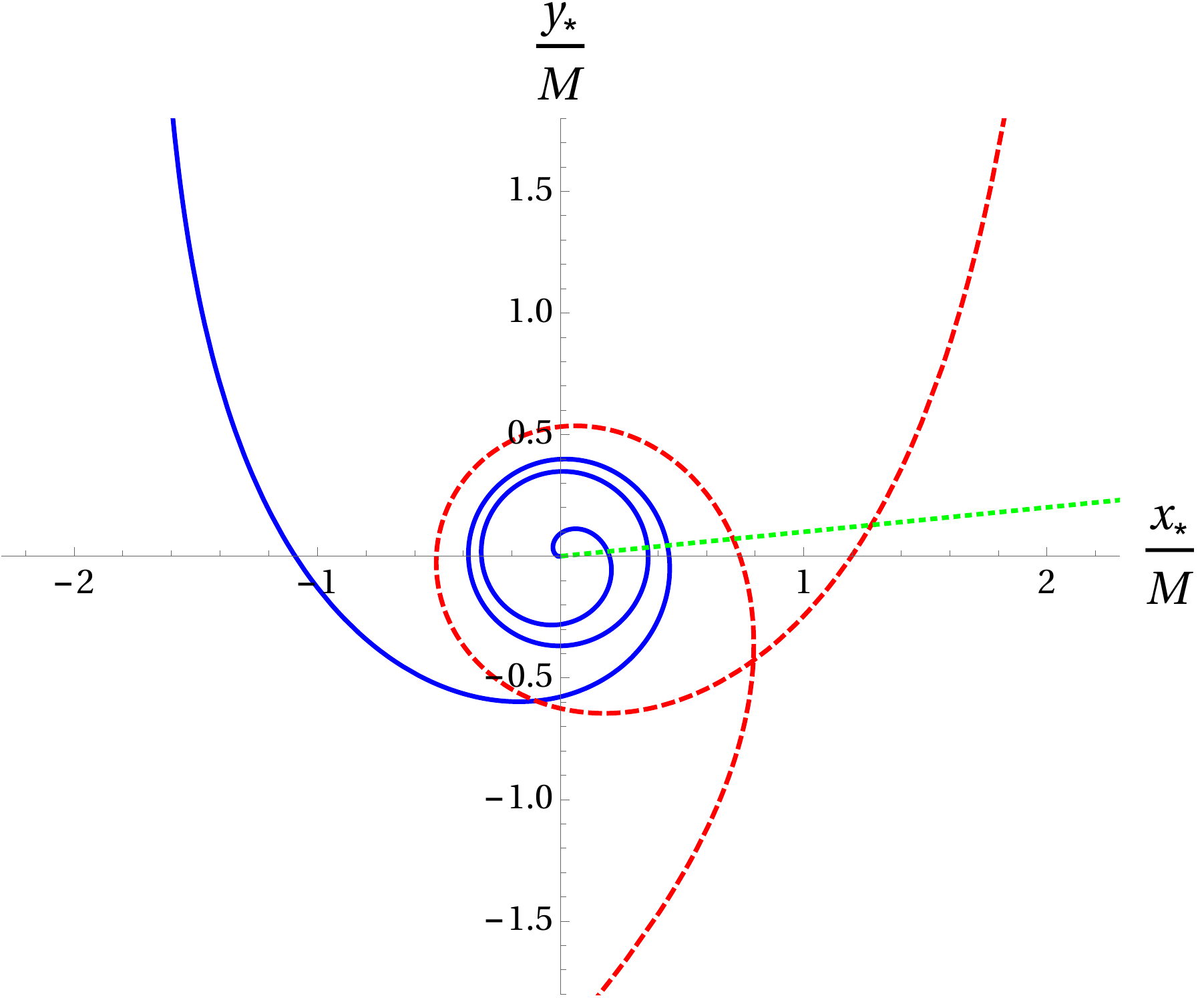}

 \vspace{.5cm}

 \includegraphics[width=8.5cm]{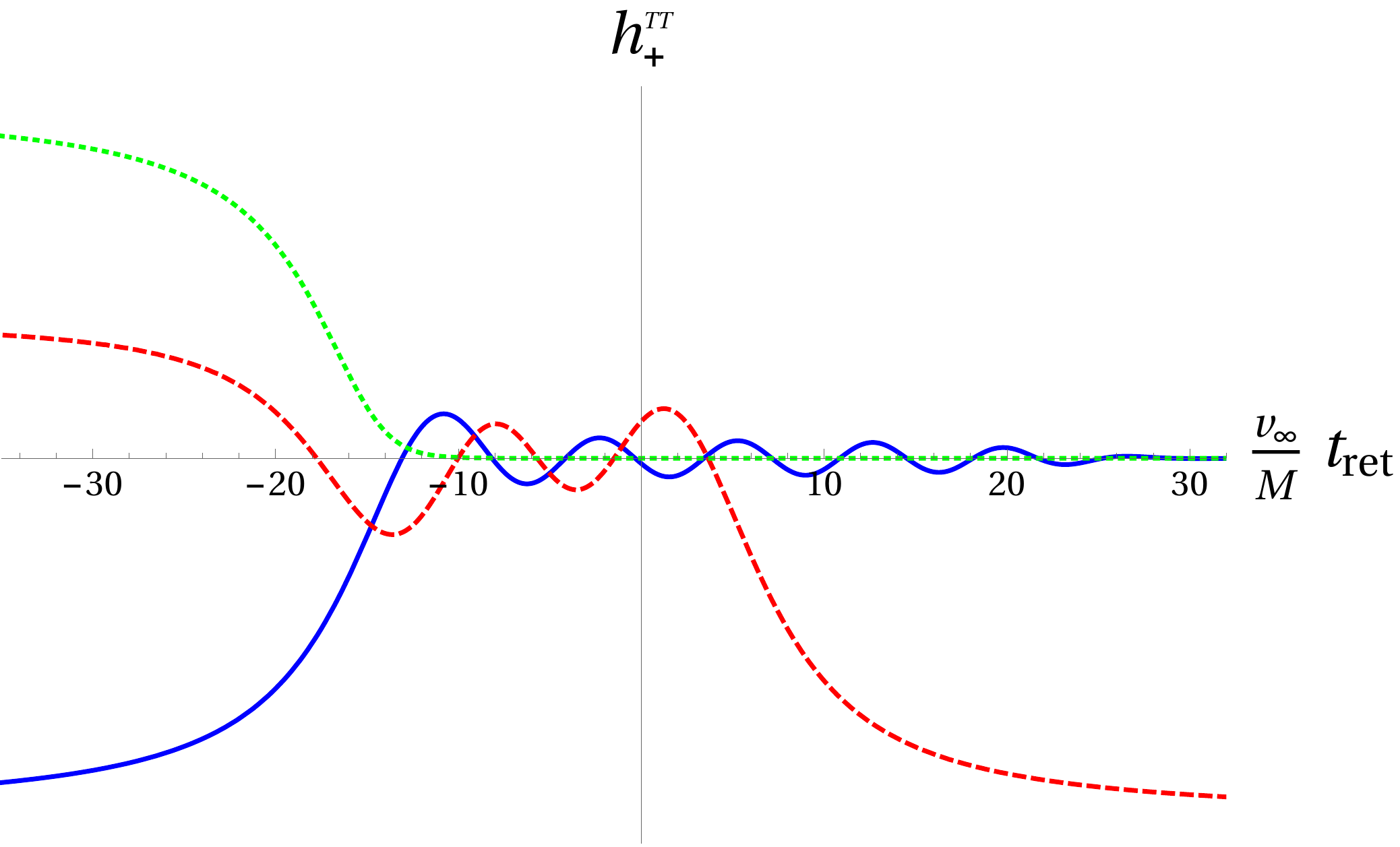}
\caption{Relative trajectories ${\bx}_*(t)$, and (north pole, $+$ polarization) gravitational waveforms $h_{+}^{TT}$, for two equal mass BHs: a head-on coalescence, in dotted green (impact parameter $b=0$); a near-critical coalescence, in solid blue ($b=2.3658\,M$); and a scattering, in dashed red ($b=2.4\, M$). For this mass ratio, $b_\mathrm{crit} \simeq 2.3660\,M$.}
\label{fg:trajsAndWaves}
\end{figure}

For simplicity and concreteness, we will henceforth display the signal arriving to an observer situated at the north pole
$\theta = 0$.\footnote{We take the north pole polarization basis defined by $\phi=0$. There is no need for such a convention in the more natural but less standard frame that drops the phase $e^{-2i\phi}$ in \eqref{circ polarization}.} The strain such an observer measures is
\bea
h^{TT} = \frac{\sqrt{2}\, \mu}{4r}  \left. \frac{d^2}{dt^2} \left(r_*^2 e^{-2i\phi_*} \right)\right|_{t_{\mathrm{ret}}} \, \hat{e}_R \, + \, \textrm{c.c.} \, .
\label{eq:SANTA}
\eea
In figure \ref{fg:trajsAndWaves} we plot the BH relative trajectory and gravitational radiation \eqref{eq:SANTA} for selected values of the impact parameter $b$.

It is instructive to work out the explicit early- and late-time limits of the waveform \eqref{eq:SANTA}. In both cases, the motion is effectively radial and the waveform is dominated by the term with no $\phi_*$ derivatives $\sim d^2 r_*^2 / dt^2 \, e^{-2i\phi_*}$. The radiation at early times is found by substituting \eqref{eq:earlyTimes} into \eqref{eq:SANTA}, giving
\bea
h^{TT}_{\mathrm{early}}= \frac{\sqrt{2}\, \mu\,v_\infty^2}{2r} \, \left[1 -\frac{3 M}{2v_\infty\,t}\right] e^{-2i\phi^{\mathrm{i}}_*}  \,\hat{e}_R +  \textrm{c.c.} \, ,
\label{eq:SANTAearly}
\eea
where $\phi^{\mathrm{i}}_*$ is the initial value of $\phi_*$.
Radiation from scattering orbits at late times $t$ is given by \eqref{eq:SANTAearly} with $v_\infty \, \to \, -v_\infty$ and $\phi^{\mathrm{i}}_* \, \to \, \phi^{\mathrm{f}}_*$ where $\phi^{\mathrm{f}}_*$ is the final value of $\phi_*$. For coalescing orbits at late times, plugging \eqref{eq:lateTimes} into \eqref{eq:SANTA} gives
\bea
h^{TT}_{\mathrm{coalescing,late}} = \frac{ 5 \, \sqrt{2} \, \mu\, C_\infty^2\,v_\infty^2}{ r} \frac{e^{-2i\phi^{\mathrm{f}}_*}}{(v_\infty\,t)^6}  \, \hat{e}_R  + \textrm{c.c.} \, .
\label{eq:SANTAlate}
\eea

The asymptotic aspects of the radiation fields \eqref{eq:SANTAearly} and \eqref{eq:SANTAlate} are in harmony with two known aspects of gravitational radiation theory. First, one sees that generically there is a piece of the late-time radiation that does not decay asymptotically -- an overall offset between the early- and late-time values of the strain:
\bea
\Delta h^{TT}= \frac{\sqrt{2}\, \mu\, v_\infty^2}{2r} \left[e^{-2i\phi^{\mathrm{f}}_*}-e^{-2i\phi^{\mathrm{i}}_*} \right]   \hat{e}_R + \textrm{c.c.} \, .
\label{eq:memoryEffect}
\eea
This is the celebrated gravitational (linear) \emph{memory effect} \cite{linear memory}. A constant strain is pure gauge; however, the difference between the asymptotically constant values of the strain is physical. This difference is the zero-frequency component of the gravitational wave, and can be observed by measuring the overall change in distance between two asymptotic observers situated at different angles on the celestial sphere.

Second, eq.~\eqref{eq:SANTAlate} shows the gravity/EM wave's decay at late times for a merger. It is known that for $t \to \infty$, a massless scalar field around a single ERN BH decays as $t^{-(2\ell+2)}$ \cite{Ori:2013iua} (and see \cite{Sela:2016qau} for a discussion of massless tensor fields). The field's decay in eq.~\eqref{eq:SANTAlate} agrees with this decay rate, given its quadrupolar, $\ell=2$ nature.

The signal (\ref{eq:quadFormula}) is exact in the sense that waveforms converge uniformly in the limit $v \to 0$. This can be seen from a power counting argument, in part reminiscent of the PN case, as follows. The effective action (\ref{eq:effectiveAction}) that determines the leading order motion is $\mathcal{O}(v^2)$. There are relativistic corrections at $\mathcal{O}(v^4)$ which are conservative and symmetric under time reversal. These lead to $\mathcal{O}(v^2)$ corrections to the leading order motion, and since the nontrivial part of a generic (noncritical) orbit lasts for a time of $\mathcal{O}(v^{-1})$, the integrated error in the trajectory, and hence in the waveform, is $\mathcal{O}(v)$. The integrated energy emitted in GWs is of $\mathcal{O}(v^5)$, and therefore its dissipative back-reaction on the trajectory is negligible.

It is important to note that the MSA breaks down in the very final stage of merger, when $r_* \sim v^2 M$. Therefore, the radiation computed here will be followed by a later signal.\footnote{This is understood to be significant at least for a certain range of $b$, $\mu$ from cosmic censorship considerations \cite{Ferrell:1987gf,FE2}.} It is interesting, however, that the decay rate $t^{-(2\ell+2)}$ appears explicitly in our calculation for moderately late times, when the BHs haven't quite merged. It would be interesting to better understand the very final stage of the coalescence.

\section*{Acknowledgements}
We are grateful to Stefanos Aretakis, Ofek Birnholtz, Gary Gibbons, Will Kelly, Chris Moore, Harvey Reall, Jorge Santos, and Uli Sperhake for useful conversations. JC is supported by the Simons Foundation through the `It from Qubit' collaboration, and thanks DAMTP for hospitality. SH is supported by the Blavatnik Postdoctoral Fellowship. NSM is partly supported by STFC grant ST/L000385/1.

\end{document}